\documentclass[twocolumn]{aastex63}

\graphicspath{{./}{figures/}}
\usepackage{pifont}
\usepackage{multirow}
\usepackage{amsmath}
\usepackage{xcolor}

\begin{document}

\title{Bumpy Superluminous Supernovae Powered by a Magnetar-star Binary Engine}

\author[0000-0002-9195-4904]{Jin-Ping Zhu}
\affiliation{School of Physics and Astronomy, Monash University, Clayton Victoria 3800, Australia; \url{jin-ping.zhu@monash.edu}}
\affiliation{OzGrav: The ARC Centre of Excellence for Gravitational Wave Discovery, Australia}

\author[0000-0002-8708-0597]{Liang-Duan Liu}
\affiliation{Institute of Astrophysics, Central China Normal University, Wuhan 430079, China}
\affiliation{Key Laboratory of Quark and Lepton Physics (Central China Normal University), Ministry of Education, Wuhan 430079, China}

\author[0000-0002-1067-1911]{Yun-Wei Yu}
\affiliation{Institute of Astrophysics, Central China Normal University, Wuhan 430079, China}
\affiliation{Key Laboratory of Quark and Lepton Physics (Central China Normal University), Ministry of Education, Wuhan 430079, China}

\author[0000-0002-6134-8946]{Ilya Mandel}
\affiliation{School of Physics and Astronomy, Monash University, Clayton Victoria 3800, Australia; \url{jin-ping.zhu@monash.edu}}
\affiliation{OzGrav: The ARC Centre of Excellence for Gravitational Wave Discovery, Australia}

\author[0000-0002-8032-8174]{Ryosuke Hirai}
\affiliation{School of Physics and Astronomy, Monash University, Clayton Victoria 3800, Australia; \url{jin-ping.zhu@monash.edu}}
\affiliation{OzGrav: The ARC Centre of Excellence for Gravitational Wave Discovery, Australia}
\affiliation{Astrophysical Big Bang Laboratory (ABBL), Cluster for Pioneering Research, RIKEN, Wako, Saitama 351-0198, Japan}

\author[0000-0002-9725-2524]{Bing Zhang}
\affiliation{Nevada Center for Astrophysics, University of Nevada, Las Vegas, NV 89154, USA}
\affiliation{Department of Physics and Astronomy, University of Nevada, Las Vegas, NV 89154, USA}

\author{Aming Chen}
\affiliation{Tsung-Dao Lee Institute, Shanghai Jiao Tong University, Shanghai 201210, China}

\begin{abstract}

Wolf-Rayet stars in close binary systems can be tidally spun up by their companions, potentially leaving behind fast-spinning highly-magnetized neutron stars, known as ``magnetars", after core collapse. These newborn magnetars can transfer rotational energy into heating and accelerating the ejecta, producing hydrogen-poor superluminous supernovae (SLSNe). In this {\em{Letter}}, we propose that the magnetar wind of the newborn magnetar could significantly evaporate its companion star, typically a main-sequence or helium star, if the binary system is not disrupted by the {abrupt mass loss and} SN kick. The subsequent heating and acceleration of the evaporated star material along with the SN ejecta by the magnetar wind can produce a post-peak bump in the SLSN lightcurve. Our model can reproduce the primary peaks and post-peak bumps of four example observed multiband SLSN lightcurves, revealing that the mass of the evaporated  material could be $\sim0.4-0.6\,M_\odot$ if the material is hydrogen-rich. {We propose that the magnetar could induce strongly enhanced evaporation from its companion star near the pericenter if the orbit of the post-SN binary is highly eccentric, ultimately generating multiple post-peak bumps in the SLSN lightcurves. This ``magnetar-star binary engine" model may offer a possible explanation for the evolution of polarization, along with the origin and velocity broadening of late-time hydrogen or helium broad spectral features observed in some bumpy SLSNe.} The diversity in the lightcurves and spectra of SLSNe may be attributed to the wide variety of companion stars and post-SN binary systems.

\end{abstract}


\section{Introduction} \label{sec:intro}

Hydrogen-poor superluminous supernovae (abbreviated to SLSNe hereafter) are a peculiar type of intrinsically bright SNe with a peak absolute magnitude $M<-21\,{\rm{mag}}$ \citep{GalYam2012,GalYam2019}, whose spectra usually lack hydrogen/helium features and commonly evolve from a hot photospheric phase with blue continua and weak absorption lines at early epochs into a cool photospheric phase resembling spectra of broad-line Type Ic SNe (SNe Ic-BL). The unusually high luminosity of SLSNe seriously challenges the traditional SN energy sources of radioactive decays of $^{56}$Ni and $^{56}$Co \citep{Nicholl2013}. Alternative scenarios are broadly grouped into two classes:  (1) circumstellar medium (CSM) interaction \citep[e.g.,][]{Chevalier2011,Chatzopoulos2012,Chatzopoulos2013} and (2) energy ejection from a central engine, such as the spin-down of a newborn millisecond magnetar \citep[e.g.,][]{Kasen2010,Woosley2010} or fallback accretion onto a black hole \citep{Dexter2013,vanPutten2017}. Detailed systematic studies further revealed that a fraction ($\sim30-80\%$) of SLSNe display significant bumps in their lightcurves \citep[e.g.,][]{Hosseinzadeh2022,Chen2023ii}. These bumps provide a potential test of SLSN models.

The lightcurve bumps can be classified into two types: (1) an early bump before the main peak \citep[e.g.,][]{Leloudas2012,Nicholl2015,Smith2016} and (2) late bumps or undulations appearing in the declining SLSN lightcurve \citep[e.g.,][]{Yan2015,Yan2017,Nicholl2016,Fiore2021,Pursiainen2022,Zhu2023,West2023}. While the pre-peak bump are typically ascribed to a shock breakout signal \citep[especially in the magnetar engine model;][]{Moriya2012,Piro2015,Kasen2016,Margalit2018,Liu2021}, the origin of the post-peak bumps is still under debate. The typical explanation is that the bumps are caused by the interaction of the SN ejecta with the CSM that is produced by the progenitor several decades before the SN explosion \citep[e.g.,][]{Yan2015,Yan2017,Nicholl2016,Wang2016,Liu2018,Li2020,Fiore2021}.  Broad H$\alpha$ or He\textsc{\,i} emission lines appear in the late-time spectra of some bumpy SLSNe \citep[e.g.,][]{Yan2015,Yan2017,Yan2020}, which seemingly support this CSM-interaction explanation. Nevertheless, it remains challenging to understand how the progenitor star can drive such violent loss of hydrogen-rich and helium-rich materials during the period just before the SN explosion. One possibility is that, if the mass of the progenitor star can be as high as $\sim(70-140)\,M_\odot$, then the progenitor could experience a pulsational pair-instability process, leading to drastic mass loss \citep{Woosley2007,Chatzopoulos2012,Woosley2017,Lin2023}. However, the event rate of such pulsational pair-instability SNe, which is limited by the stringent mass requirement for the progenitors, could be too low to explain the relatively large number of observed SLSNe. Moreover, the high mass of the SN ejecta could be in tension with the rapidly evolving lightcurves of a remarkable fraction of SLSNe.

Alternatively, in the magnetar engine model, the lightcurve bumps can in principle be attributed to the intermittent violent energy releases from the magnetar \citep{Yu2017Flare,Dong2023} or the variable thermalization efficiency of the injected energy \citep{Vurm2021,Moriya2022}, while the general shape of the SLSN lightcurves can usually successfully be modeled by the magnetar spin-down whose temporal behavior scales as $L\propto{t}^{-2}$ \citep[e.g.,][]{Inserra2013,Yu2017,Liu2017,Nicholl2017,Blanchard2020}. However, the specific mechanism that would cause the significant variation in the energy released by the magnetar is uncertain, though it would presumably be related to the progenitor star and magnetar formation physics.

The classical model suggested SLSN progenitors could arise from stars that can experience quasi-chemically homogeneous evolution \citep[e.g.,][]{
AguileraDena2018,Song2023,Ghodla2023}, directly evolving into fast-spinning and compact Wolf-Rayet stars. On the other hand,  \citet{Fuller2022,Hu2023} proposed that the SLSN magnetars could originate from the core collapse of Wolf-Rayet stars that were tidally spun up by their companions in very compact binaries formed via dynamically unstable (common-envelope) or stable mass transfer, which was strongly hinted by the universal energy-mass correlation discovered by \cite{Yu2017} and \cite{LiuJF2022} . The detailed simulation by \cite{Hu2023} clearly exhibited the pretty consistence of the binary origin of the fast-spinning magnetars with the observational correlation among SLSNe, gamma-ray bursts, SNe Ic-BL, and fast blue optical transients as well as their event rates. 

For such a close binary progenitor, the newborn magnetar could remain bound to the companion star. In other words, the central engine of the SLSN could be a binary consisting of a magnetar and a main-sequence (MS) or helium star, rather than a solitary magnetar. Then, a powerful {magnetar wind (MW)} can evaporate the MS/helium star companion and eject outflows\footnote{Detailed SN explosion models do not unambiguously connect rapidly rotating progenitors with magnetar formation \citep[e.g.,][]{TorresForne2016,MuellerVarma2020}.  Nor is there clear observational evidence for rapid rotation or pulsar wind nebulae in SN remnants hosting magnetars \citep{Vink2008}.  Therefore, our operating assumption that rapidly rotating magnetars born in close binaries can drive energetic and long-duration MWs has not yet been definitively validated, either theoretically or observationally.}, which may be intermittent if the orbit is highly eccentric.  

In this {\em Letter}, we assess the basic properties of this evaporative outflow and investigate the effect of the outflow on the SLSN lightcurve.  In the proposed model, evaporated material creates a broad torus in the orbital plane of the binary, which delays diffusion {due to the increased amount of material along the equatorial direction and the higher opacity of the evaporated material compared to that of the SN ejecta}. The primary lightcurve peak then corresponds to the time when energy can diffuse in directions where there is little evaporated material, while later bumps correspond to diffusion through the combination of SN ejecta and evaporated material.  Meanwhile, the broad-line H$\alpha$ or He\textsc{\,i} features can be naturally produced by the hydrogen-rich or helium-rich outflows from the MS/helium companion, which can finally be detected as the SN ejecta become transparent at late times. {The {\em Letter} is organized as follows. The magnetar-star engine is introduced in Section \ref{sec:Model}. We present an analytical lightcurve model for SNe powered by the magnetar-star engine and provide lightcurve fits for four example bumpy SLSNe in Section \ref{sec:lightcurve}. The summary and discussion are in Section \ref{sec:DiscussionConclusion}.}

\section{Magnetar-star Binary Engine} \label{sec:Model}

\begin{figure}[t]
\centering
\includegraphics[width=1\linewidth]{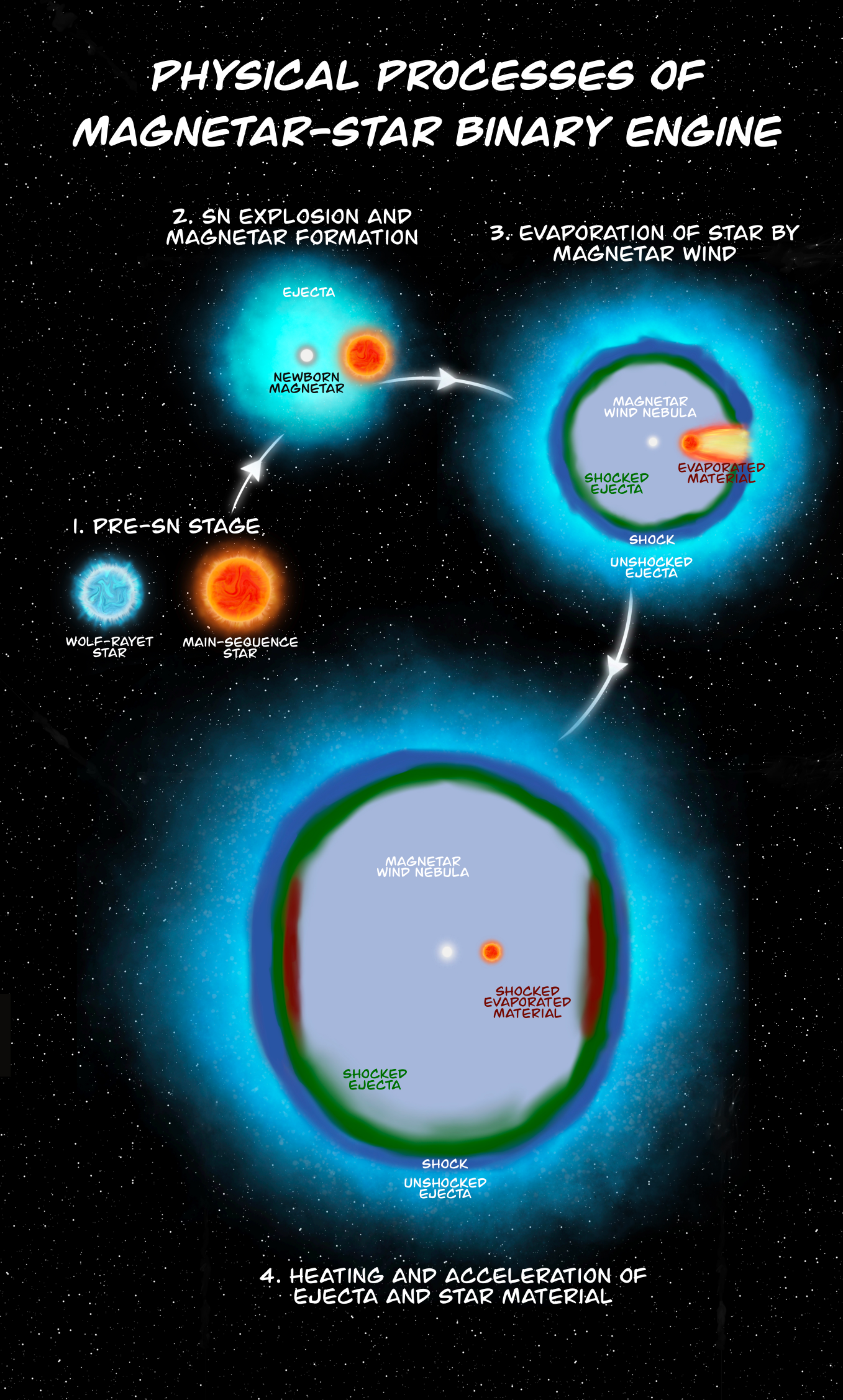}
\caption{An illustration of the physical stages of the magnetar-star binary engine.}
\label{fig:Cartoon}
\end{figure}

In Figure \ref{fig:Cartoon}, we present a schematic picture to describe the stages of the magnetar-star binary engine: (1) the pre-SN stage, (2) the SN explosion and magnetar formation, (3) the evaporation of the star by the MW, and (4) heating and acceleration of the approximately spherical SN ejecta and the torus-like evaporated material. In the following sections, we will describe these processes and explore the origin of late-time SLSN lightcurve bumps caused by the magnetar-star binary engine.

\subsection{Evaporation of a Companion} \label{sec:evaporation}

After the SN explosion of a Wolf-Rayet star in a close binary, a rapidly rotating magnetar may form. This magnetar loses its rotation energy via the magnetic dipole spin-down process with a luminosity of \citep{Ostriker1971}
\begin{equation}
    L_{\rm sd} = \frac{L_{\rm sd,i}}{(1 + t/t_{\rm sd})^2},
\end{equation}
where the initial luminosity is $L_{\rm sd,i} \simeq E_{\rm rot}/t_{\rm sd} = B_{\rm p}^2R_{\rm mag}^6\Omega_{\rm i}^4/6c^3\simeq 10^{47}B_{\rm p,14}^2P_{\rm i,-3}^{-4}\,{\rm erg}\,{\rm s}^{-1}$ and the spin-down timescale is $t_{\rm sd}=3c^3I_{\rm mag}/B_{\rm p}^2R_{\rm mag}^6\Omega_{\rm i}^2\simeq2\times10^5B_{\rm p,14}^{-2}P_{\rm i,-3}^2\,{\rm s}$ where $B_{\rm p}$ is the newborn magnetar's polar magnetic field strength, $E_{\rm rot}=I_{\rm mag}\Omega_{\rm i}^2/2$ is the rotational energy, $\Omega_{\rm i}$ and $P_{\rm i}=2\pi/\Omega_{\rm i}$ are the initial angular frequency and spin period, $I_{\rm mag}\simeq10^{45}{\rm g}\,{\rm cm}^2$ is the moment of inertia,  $R_{\rm mag}\simeq10^6{\rm cm}$ is the {magnetar's} radius, and $c$ is the speed of light. Hereafter, we use the conventional notation $Q_{x} \equiv Q / 10^x$ in cgs units.

If the post-SN close-orbit binary system is not disrupted by the {abrupt mass loss and} SN kick, the high-pressure MW from the newborn magnetar can evaporate the companion star with the entire process occurring within the enclosure of the SN ejecta. By assuming a circular orbit of the post-SN binary system, the evaporation rate of the companion star can be given by \citep{vandenHeuvel1988}
\begin{equation}
\begin{split}
    \dot{M}_{\rm ev} &= f\left(\frac{R_\star}{a}\right)^2\frac{L_{\rm sd}}{2v_{\rm esc}^2} \\
    &=0.1\,f_{-1}M_{\star,4\odot}^{-1}R_{\star,3\odot}^3a_{10\odot}^{-2}L_{\rm sd,i,45.5}\,M_\odot\,{\rm d}^{-1}, \\
\end{split}
\end{equation}
where $f$ is the fraction of the incident wind energy that can be transferred to the companion star, $M_{\star,4\odot} = M_\star/(4\,M_\odot)$ is the companion's mass, $R_{\star,3\odot}=R_\star/(3\,R_\odot)$ is the companion's radius, $a_{10\odot}=a/(10\,R_\odot)$ is the orbital radius of the post-SN orbit, and $v_{\rm esc} = \sqrt{2GM_\star/R_\star} = 710\,M_{\star,4\odot}^{1/2}R_{\star,3\odot}^{-1/2}\,{\rm km}\,{\rm s}^{-1}$ is the escape velocity with gravitational constant $G$. The time for completely evaporating the star by the MW can be thus estimated as
\begin{equation}
\label{equ:evaporatedTime}
    t_{\rm ev}\approx\frac{M_\star}{\dot{M}_{\rm ev}} = 33\,f^{-1}_{-1}M_{\star,4\odot}^2R_{\star,3\odot}^{-3}a_{10\odot}^2L_{\rm sd,i,45.5}^{-1}\,{\rm d}.
\end{equation}
We can then estimate the mass of the evaporated material as
\begin{equation}
\begin{split}
    {M}_{\rm ev}&\approx \dot{M}_{\rm ev} \min(t_{\rm ev},t_{\rm sd}) \\
    &\approx \min(M_\star,f_{-1}M_{\star,4\odot}^{-1}R_{\star,3\odot}^3a_{10\odot}^{-2}E_{\rm rot,51.5}\,M_\odot).
\end{split}
\end{equation}
Therefore, complete evaporation of companion stars in a fraction of systems can be achievable within $t_{\rm ev}\lesssim t_{\rm sd}$ if the MWs are powerful, the companion stars are not too massive, the post-SN systems are very close, and the energy transfer efficiency is very high; otherwise, the stars can be partially evaporated by the newborn magnetars\footnote{In addition to the evaporation of the stars by the newborn magnetars, SN explosions can inject energy into the companion stars \citep{Hirai2018,Ogata2021,Hirai2023}, causing them to inflate and thus facilitating their evaporation by the magnetars.}. 

Tides are expected to efficiently spin up Wolf-Rayet stars in close binaries with an orbital period of $\lesssim2\,{\rm d}$ \citep[e.g.,][]{Qin2018,FullerMa2019,Hu2023}, causing them to leave behind rapidly spinning magnetars that drive SLSN explosions. This corresponds to a pre-SN orbital radius of $a\lesssim13\,R_\odot$ for two stars with an equal mass of $4\,M_\odot$. Previous statistical studies \citep[e.g.,][]{Yu2017,Liu2017,Nicholl2017,Hosseinzadeh2022} showed that the typical values of $L_{\rm sd,i}$, $t_{\rm sd}$ and ejecta mass $M_{\rm ej}$ for the observed SLSN population are $\sim3\times10^{45}{\rm erg}\,{\rm s}^{-1}$, $\sim10\,{\rm d}$, and  $\sim4\,M_\odot$, respectively. For these representative values, one can expect that the mass of the evaporated material $M_{\rm ev}$ should generally exceed a few $0.1\,M_\odot$ by assuming that the pre-SN and post-SN orbital radii are similar and the companion mass is similar to the ejecta mass.

The evaporated outflow will be initially accelerated outward along the line extending from the magnetar to the star due to the acceleration by the MW. Defining the orbital plane of the magnetar-star binary as the equatorial plane, the evaporated material would ultimately form a torus-like structure \citep[e.g.,][]{Yu2019} distributed around this plane. We denote the half-opening angle of evaporated material in the latitudinal direction with $\theta_{\rm ev}$.  The torus will occupy a fraction $f_{\rm ev}= \sin\theta_{\rm ev}$ of the sky as viewed from the magnetar.  The angular radius of the companion as viewed from the magnetar is $\tan(\theta_\star) = R_\star/a$; since $a\lesssim13\,R_\odot$ for two $4\,M_\odot$ stars, the torus could  typically occupy $f_{\rm ev}\gtrsim0.23\,{R}_{\star,3\odot}a^{-1}_{13\odot}$ of the sky if $\theta_{\rm ev} \approx \theta_\star$.

\subsection{Late-time Bump}

In the polar direction of a magnetar-star close-orbit binary system, the newborn magnetar can directly heat and accelerate the ejecta to drive magnetar-powered SLSNe without being blocked by the evaporated material. Its peak time can be determined by the effective diffusion time of
\begin{equation}
\label{equ:DiffusionTime1}
    t_{\rm d,p} \approx \left( \frac{3\kappa_{\rm ej}M_{\rm ej}}{4\pi v_{\rm ej,p}c} \right)^{1/2}\approx 31\,M_{\rm ej,4\odot}^{3/4}L_{\rm sd,i,45.5}^{-1/2}t_{\rm sd,6}^{-1/2}\,{\rm d},
\end{equation}
where $\kappa_{\rm ej}=0.1\,{\rm cm}^2\,{\rm g}^{-1}$ is the opacity for ejecta dominated by carbon and oxygen \citep[e.g.,][]{Inserra2013,Nicholl2015}, and $v_{\rm ej,p}\approx\sqrt{2(E_{\rm rot} + E_{\rm SN})/M_{\rm ej}} \approx \sqrt{2 E_{\rm rot}/M_{\rm ej}}$.  Here, we assume that for a magnetar-powered SLSN, the energy extracted from the magnetar spin-down dominates over the core-collapse SN explosion energy, $E_{\rm rot}\gg{E}_{\rm SN} \approx 10^{51}\,{\rm erg}$.  Based on the  \citet{Arnett1982} law, the luminosity at peak can be approximated as the instantaneous spin-down luminosity at the peak time, i.e., $L_{\rm mag,p}^{\rm peak} \approx (1-f_{\rm ev}) L_{\rm sd}(t_{\rm d,p}) \approx 3.1\times10^{44} \left(\frac{1-f_{\rm ev}}{0.7}\right) M_{\rm ej,4\odot}^{-3/2}L_{\rm sd,i,45.5}^{2}t_{\rm sd,6}^{3} \,{\rm erg}\,{\rm s}^{-1}$ if $t_{\rm sd} \ll t_{\rm d,p}$.

Near the equatorial plane of the system, the magnetar needs to heat and accelerate both the ejecta and evaporated material.  Because the evaporated material initially blocks the SN ejecta, MW energy is first deposited into this material, accelerating it until it catches up with the ejecta.  At this stage, the  evaporated material, which previously had a broad velocity distribution, would be swept into a thin shell located within the inner part of the ejecta by the MW.  After a time of order the diffusion time through the combined SN ejecta and evaporated material, magnetar-powered emission can emerge through the torus near the equatorial plane. Similar to Equation (\ref{equ:DiffusionTime1}), the peak time of the equatorial magnetar-powered emission is
\begin{equation}
    t_{\rm d,e} \approx \left[ \frac{3(\kappa_{\rm ej}M_{\rm ej} + \kappa_{\rm ev}M_{\rm ev}/f_{\rm ev})}{4\pi{v}_{\rm ej,e}c}\right]^{1/2},
\end{equation}
where $v_{\rm ej,e}=\sqrt{2(E_{\rm rot}+E_{\rm SN})/(M_{\rm ej}+M_{\rm ev}/f_{\rm ev})}$ is the velocity of the equatorial ejecta and evaporated material. Since most of the stars accompanying SLSN progenitors could be MS stars, our calculations only consider the evaporated material to be hydrogen-rich. The opacity of the fully-ionized hydrogen-rich material is set to be $\kappa_{\rm ev}=0.33\,{\rm cm}^2\,{\rm g}^{-1}$ \citep{Moriya2011,Chatzopoulos2012}. In Section \ref{sec:evaporation}, we have estimated the mass of the evaporated material $M_{\rm ev}$, which should generally exceed a few $0.1\,M_\odot$. Although $M_{\rm ev}$ is much lower than $M_{\rm ej}$, the peak time of the equatorial magnetar-powered emission can be significantly delayed because the evaporated material is concentrated near the equatorial plane and $\kappa_{\rm ev} > \kappa_{\rm ej}$. For example, if $M_{\rm ej}=4\,M_\odot$, $E_{\rm rot}=3\times10^{51}{\rm erg}$, $M_{\rm ev}=0.5\,M_\odot$, and $f_{\rm ev}=0.3$, one can obtain $t_{\rm d,e} \approx 52\,{\rm d}$, longer than $t_{\rm d,p}\approx31\,{\rm d}$ in Equation (\ref{equ:DiffusionTime1}). The peak radiated luminosity of the equatorial magnetar-powered emission is given by $L_{\rm mag,e}^{\rm peak} \approx f_{\rm ev} L_{\rm sd}(t_{\rm d,e})$, or a fraction $f_{\rm ev}/(1-f_{\rm ev})\gtrsim 30\%$ of the radiated luminosity of the polar magnetar-powered emission at $t_{\rm d,e}$. If observers treat the early polar magnetar-powered emission as the peak SLSN emission, the late-time equatorial magnetar-powered emission would be interpreted as a bump in the lightcurve.

\section{Lightcurve} \label{sec:lightcurve}

\subsection{Models}

For the magnetar-powered from the polar direction, we calculate the bolometric luminosity by including the gamma-ray leakage effect in the standard SN diffusion equation as \citep{Arnett1982,Wang2015}
\begin{equation}
\begin{split}
\label{equ:magnetarLightcurve}
    L_{\rm mag,p}&(t) = (1-f_{\rm ev})e^{-(t/t_{\rm d,p})^2}\\
    &\int^{t}_02L_{\rm sd}(t')\frac{t'}{t_{\rm d,p}}e^{(t'/t_{\rm d,p})^2}\left(1-e^{-A_{\rm p}t'^{-2}}\right)\frac{dt'}{t_{\rm d,p}},
\end{split}
\end{equation}
where the factor $A_{\rm p} = 3\kappa_{\gamma,\rm p}M_{\rm ej}/4\pi v_{\rm ej,p}^2$ accounts for  gamma-ray leakage and $\kappa_{\gamma,\rm p}$ is the effective gamma ray opacity for the polar ejecta. Similar to Equation (\ref{equ:magnetarLightcurve}), the magnetar-powered bolometric lightcurve from the equatorial direction can be written as
\begin{equation}
\begin{split}
    L&_{\rm mag,e}(t) = f_{\rm ev}e^{-(t/t_{\rm d,e})^2}\\
    &\int^{t}_{t_{\rm delay}}2L_{\rm sd}(t')\frac{t'}{t_{\rm d,e}}e^{(t'/t_{\rm d,e})^2}(1-e^{-A_{\rm e}t'^{-2}})\frac{dt'}{t_{\rm d,e}},
\end{split}
\end{equation}
where the factor $A_{\rm e}$ is given by $A_{\rm e} = 3\kappa_{\gamma,\rm e}(M_{\rm ej}+M_{\rm ev}/f_{\rm ev})/4\pi v_{\rm ej,e}^2$ and $\kappa_{\gamma,\rm e}$ is the effective gamma ray opacity for the equatorial ejecta and evaporated material.  The delay time $t_{\rm delay}$ reflects the time it takes for the evaporation of the companion by the MW and for {the last of} the evaporated material {(i.e., the innermost evaporated material)} to catch up to the ejecta. {The delay time can be estimated by equating the distance travelled by the ejecta to that of the innermost evaporated material, i.e., $v_{\rm ej,e} t_{\rm delay} \approx {v}_{\rm ev}(t_{\rm delay}-t_{\rm sd})$, so that  $t_{\rm delay}\approx f_{\rm delay}t_{\rm sd}$, where the factor is $f_{\rm delay} = v_{\rm ev}/(v_{\rm ev} - v_{\rm ej,e})$ with the velocity of the innermost evaporated material being $v_{\rm ev}\approx\sqrt{2f_{\rm ev}L_{\rm sd}/\dot{M}_{\rm ev}}\approx\sqrt{2f_{\rm ev}E_{\rm rot} / M_{\rm ev}}$. For example, if $E_{\rm SN}=10^{51}{\rm erg}$, $M_{\rm ej}=4\,M_\odot$, $t_{\rm sd}=10\,{\rm d}$, $E_{\rm rot} = 3\times10^{51}{\rm erg}$, $M_{\rm ev}=0.5\,M_\odot$, and $f_{\rm ev} = 0.3$, the factor is $f_{\rm delay}\approx3$ and the delay time can be $t_{\rm delay}\approx30\,{\rm d}$. $t_{\rm delay}$ can be significantly increased by more powerful core-collapse SNe and by relatively slowly-rotating magnetars with lower magnetic field strengths (i.e., lower $L_{\rm sd,0}$ and longer $t_{\rm sd}$).} Some uncertain processes, such as thermal energy transformation and thermalization heating {for high-energy photons from MW}, {can also contribute to the delay time at the level of $\sim20-40\,{\rm d}$ \citep[e.g.,][]{Liu2021}.} {The main physical mechanisms that could be important in determining these uncertain processes are the pair production via $\gamma$-$\gamma$ interaction \citep{Kotera2013} and the Bethe-Heiter process in the evaporated material and ejecta \citep{Murase2015,Kashiyama2016}.} In our fits, we treat $t_{\rm delay}$ as a free parameter due to these uncertainties. {One can expect that the entire delay time could range from several tens of days to over a hundred days.}

Furthermore, emission powered by the radioactive decay of $^{56}$Ni and $^{56}$Co, as well as the emission powered by the interaction between the SN ejecta and accelerated evaporated material, can also contribute to SLSN lightcurves. However, as discussed in Appendix \ref{sec:OtherEmission}, we find that these two sources of emission are always subdominant to the magnetar-powered emission and hence we ignore their contribution to the SLSN lightcurves. Therefore, the total bolometric luminosity is $L_{\rm tot}(t) = L_{\rm mag,p}(t) + L_{\rm mag,e}(t)$.

\subsection{Lightcurve Fitting}
In this section, we apply our model to multi-band SLSN lightcurves with one post-peak bump.  The origin of multiple post-peak bumps in some SLSN lightcurves will be discussed in Section \ref{sec:DiscussionConclusion}. 

From the literature, we select SLSNe with lightcurves that are well sampled with a distinct primary peak for both the rising and declining phases in at least two filters. Furthermore, our criterion requires the late-time lightcurves of our selected SLSNe to deviate from the smooth decline trend of the primary peak, displaying one post-peak bump that contains a complete rising and declining structure. Our sample includes four SLSNe whose basic information is listed in Table \ref{tab:Information}; we assume the redshifts are known perfectly when converting them into luminosity distances for our fits. { Here, a cosmology of $H_0=67.4\,{\rm km}\,{\rm s}^{-1}\,{\rm Mpc}^{-3}$, $\Omega_{\rm m}=0.315$, and $\Omega_\Lambda = 0.685$ \citep{Planck2020} is applied.}

\begin{deluxetable}{lcc}[tb]
\tabletypesize{\footnotesize}
\tablecaption{Basic information for selected well-sampled SLSNe} \label{tab:Information}
\tablehead{
\colhead{Name} &
\colhead{Redshift} &
\colhead{Reference}
}
\startdata
PTF10hgi & 0.100 & \cite{Inserra2013,DeCia2018}\\
PS1-12cil & 0.32 & \cite{Lunnan2018}\\
SN2018kyt & 0.1080 & \cite{Yan2020,Chen2023i} \\
SN2019stc & 0.1178 & \cite{Gomez2021,Chen2023i}
\enddata
\end{deluxetable}

\begin{figure*}[t]
\centering
\includegraphics[width=0.495\linewidth, trim = 85 36 94 33, clip]{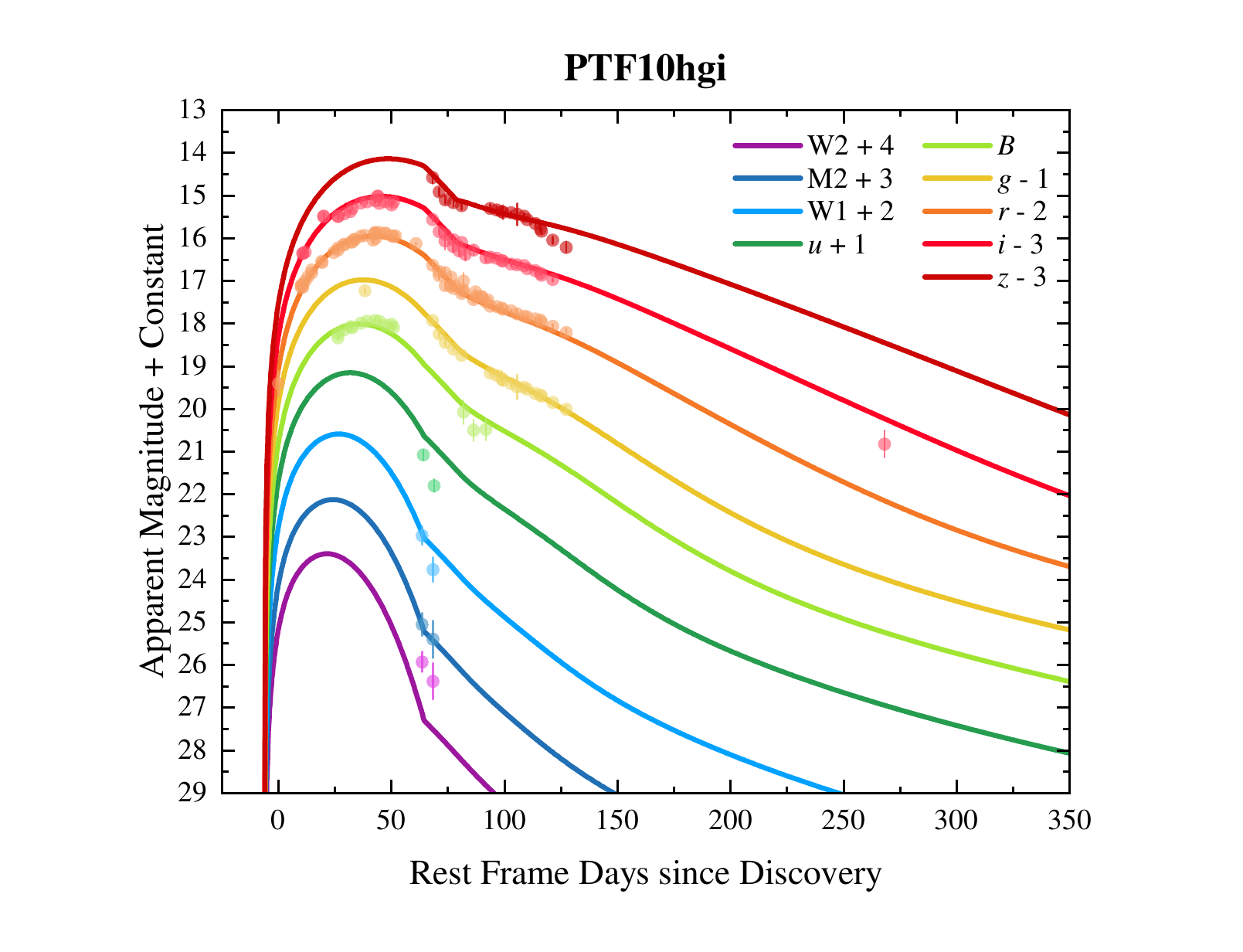}
\includegraphics[width=0.495\linewidth, trim = 85 36 94 33, clip]{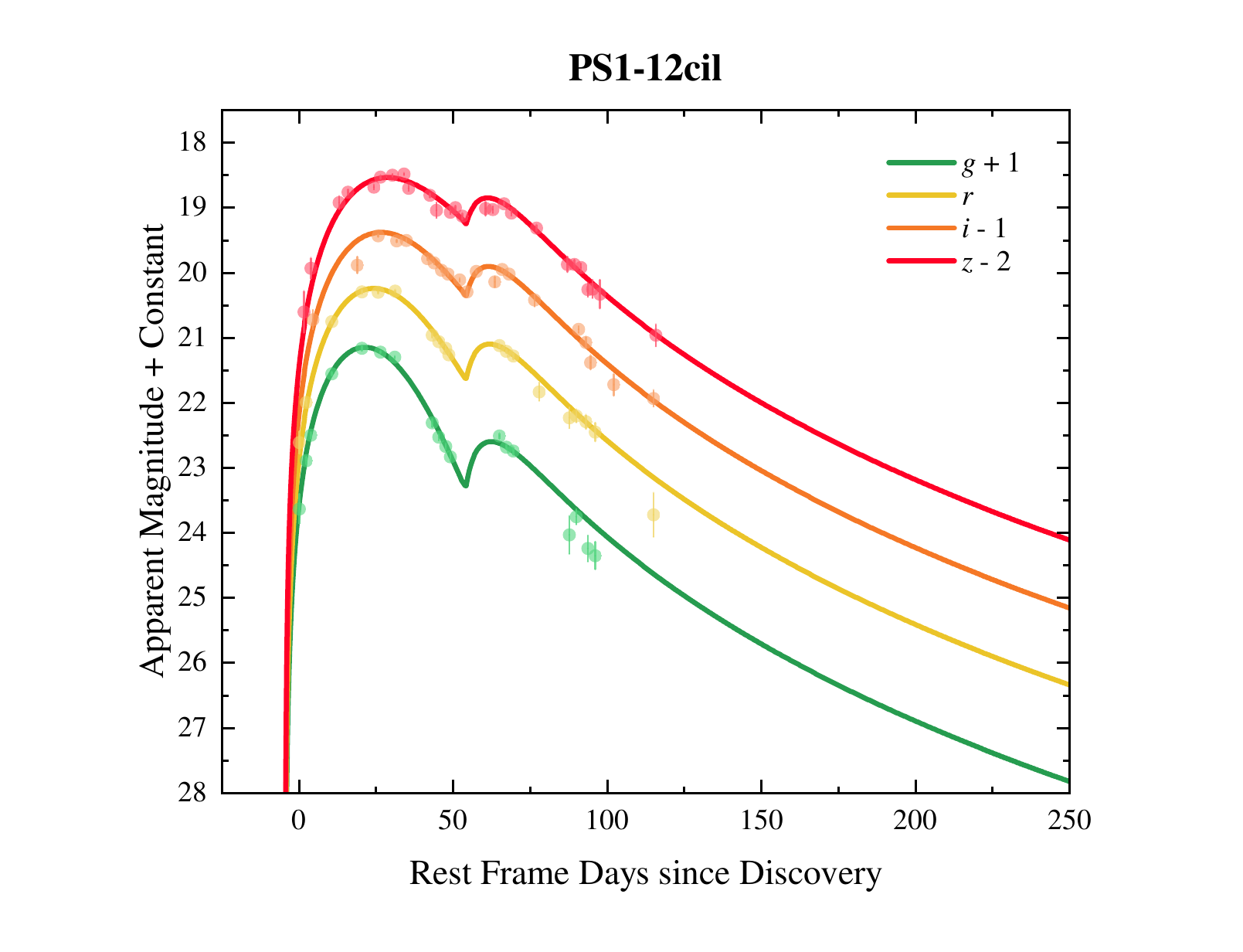}
\includegraphics[width=0.495\linewidth, trim = 85 36 94 23, clip]{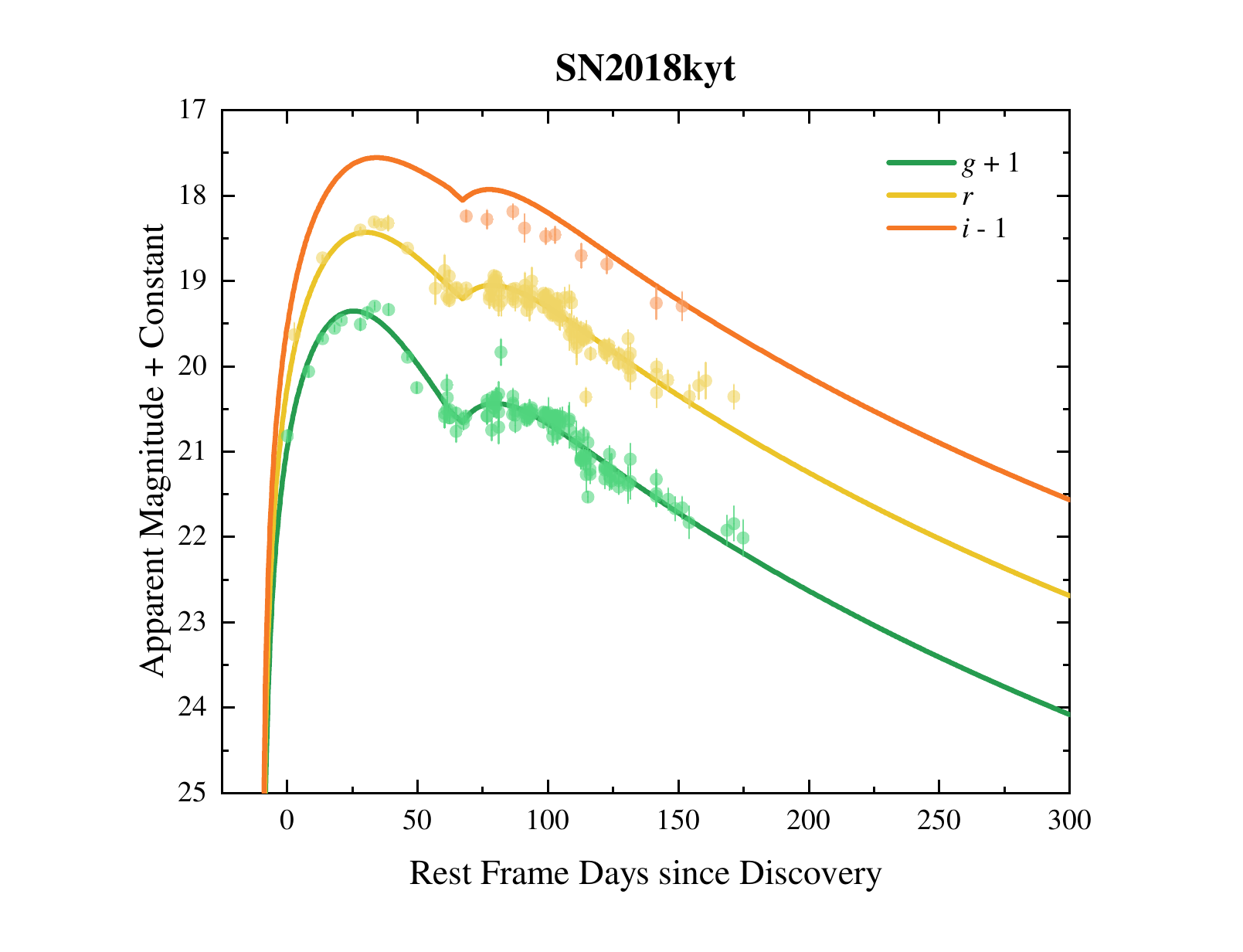}
\includegraphics[width=0.495\linewidth, trim = 85 36 94 23, clip]{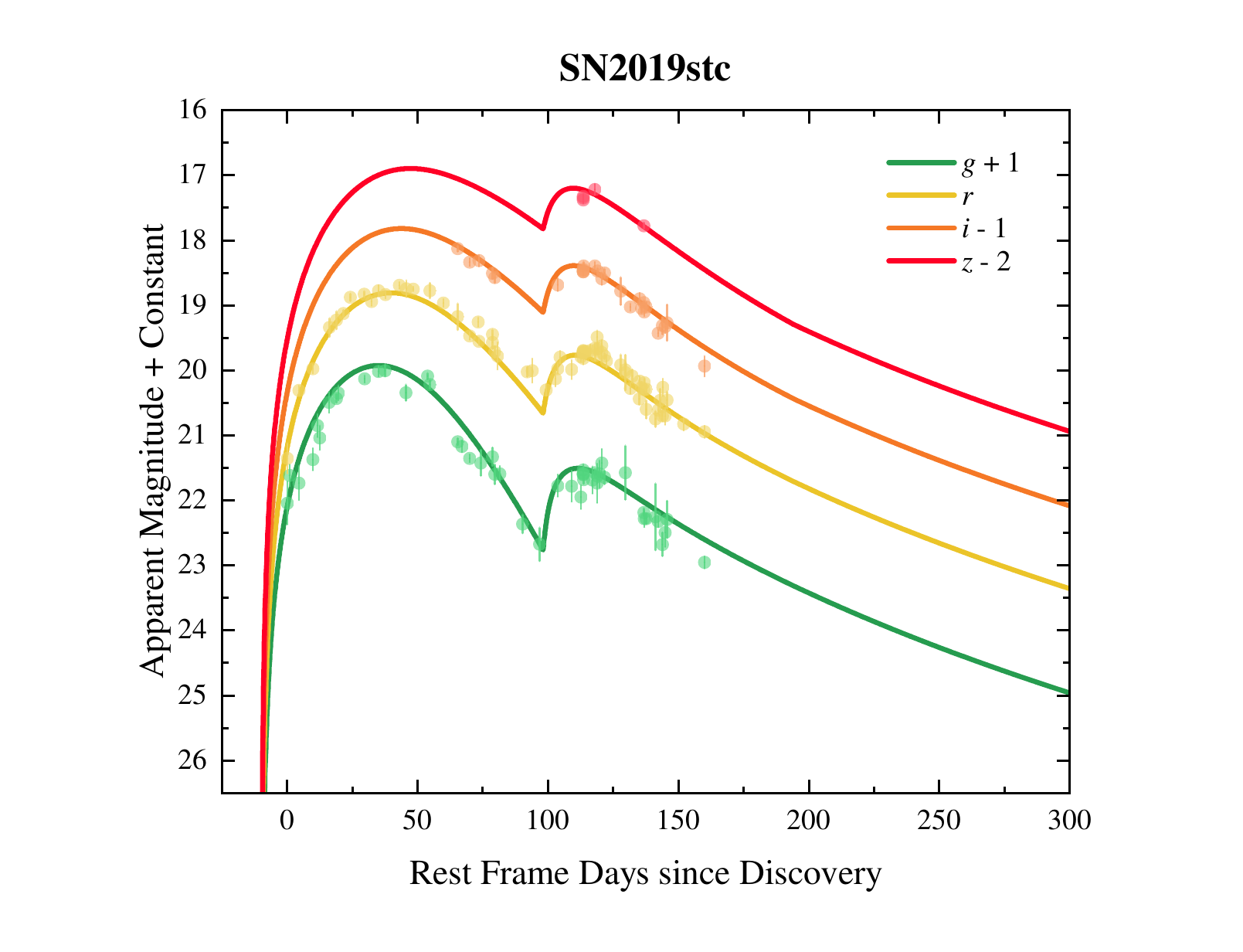}
\caption{Best fits (solid lines) to multi-wavelength lightcurves of selected bumpy SLSNe (dots with error bars) in our magnetar-star binary engine model. The label in each panel shows the different bands represented by different colors.}
\label{fig:Fitting_Results}
\end{figure*}

\begin{deluxetable*}{c|cc|rrrr}[htpb]
\tabletypesize{\footnotesize}
\tablecaption{Fitting parameter priors and posteriors for some well-sampled bumpy SLSNe} \label{tab:PriorsAndResults}
\tablehead{\multirow{2}{*}{Parameter} &
\multicolumn{2}{c}{Prior} & \multicolumn{4}{c}{Posterior}  \\
& \colhead{Distribution} &
\colhead{Bounds} &
\colhead{PTF10hgi} &
\colhead{PS1-12cil} &
\colhead{SN2018kyt} &
\colhead{SN2019stc}
}
\startdata
$P_{\rm i}/{\rm ms}$ & Flat & $[0.7 , 20]$ & $2.94^{0.08}_{0.09}$ & $1.99^{0.17}_{0.15}$ & $4.15^{0.12}_{0.15}$ & $2.25^{0.33}_{0.28}$ \\
$B_{\rm p}/10^{14}{\rm G}$ & Log-flat & $[0.1,10]$ & $1.15^{0.04}_{0.04}$ & $1.04^{0.15}_{0.13}$ & $1.44^{0.03}_{0.04}$ & $0.64^{0.11}_{0.10}$ \\
$M_{\rm ej}/M_\odot$ & Flat & $[0.1 , 30]$ & $6.10^{0.26}_{0.18}$ & $2.65^{0.23}_{0.14}$ & $2.82^{0.15}_{0.14}$ & $4.86^{0.44}_{0.44}$ \\
$M_{\rm ev}/M_\odot$ & Flat & $[0.1 , 30]$ & $0.43^{0.10}_{0.08}$ & $0.44^{0.10}_{0.10}$ & $0.60^{0.08}_{0.07}$ & $0.49^{0.25}_{0.13}$ \\
$f_{\rm ev}$ & Flat & $[0.01 , 0.99]$ & $0.24^{0.01}_{0.01}$ & $0.66^{0.04}_{0.04}$ & $0.48^{0.02}_{0.01}$ & $0.66^{0.06}_{0.04}$ \\
$T_{\rm fl,p}/10^3{\rm K}$ & Log-flat & $[1 , 20]$ & $7.69^{0.24}_{0.20}$ & $6.72^{0.22}_{0.35}$ & $6.46^{0.14}_{0.12}$ & $2.08^{0.62}_{0.70}$ \\
$T_{\rm fl,e}/10^3{\rm K}$ & Log-flat & $[1 , 20]$ & $1.37^{0.57}_{0.26}$ & $8.48^{0.31}_{0.26}$ & $7.53^{0.27}_{0.24}$ & $7.24^{0.27}_{0.28}$ \\
$\log_{10}(\kappa_{\gamma,{\rm p}}/{\rm cm}^2{\rm g}^{-1})$ & Flat & $[-3 , 3]$ & $-2.49^{0.07}_{0.08}$ & $-1.41^{0.08}_{0.06}$ & $-0.72^{0.10}_{0.09}$ & $-1.69^{0.09}_{0.09}$ \\
$\log_{10}(\kappa_{\gamma,{\rm e}}/{\rm cm}^2{\rm g}^{-1})$ & Flat & $[-3 , 3]$ & $1.00^{0.94}_{0.55}$ & $-1.53^{0.15}_{0.14}$ & $-1.42^{0.10}_{0.09}$ & $-1.66^{0.15}_{0.17}$ \\
$A_V/{\rm mag}$ & Flat & $[0 , 1]$ & $0.32^{0.03}_{0.04}$ & $0.15^{0.03}_{0.03}$ & $0.01^{0.02}_{0.01}$ & $0.80^{0.07}_{0.12}$ \\
$t_{\rm delay}/{\rm d}$ & Flat & $[0 , 300]$ & $83.52^{0.93}_{0.27}$ & $58.34^{0.46}_{0.33}$ & $76.85^{0.40}_{0.25}$ & $107.28^{0.74}_{0.43}$ \\
$t_{\rm shift}/{\rm d}$ & Flat & $[0 , 300]$ & $6.18^{0.16}_{0.21}$ & $4.48^{0.31}_{0.38}$ & $6.03^{0.45}_{0.18}$ & $9.11^{0.59}_{0.30}$
\enddata
\tablecomments{We note that $T_{\rm fl,e}$ of PTF10hgi might be imprecisely constrained. This is due to the main peak not showing a clear declining lightcurve, resulting in the posterior of $T_{\rm fl,e}$ being similar to the prior.   }
\end{deluxetable*}

In order to calculate the monochromatic luminosity of SLSNe, we define the photosphere temperature of magnetar-powered emission from the polar SN ejecta with bolometric luminosity $L_{\rm mag,p}$ as
\begin{equation}
\label{equ:Temperature}
    T_{\rm ph,p} =\max\left[\left(\frac{L_{\rm mag,p}}{4\pi(1-f_{\rm ev})\sigma_{\rm SB} v_{\rm ph,p}^2 t^2}\right)^{1/4},T_{\rm fl,p}\right],
\end{equation}
where $\sigma_{\rm SB}$ is the Stefan-Boltzmann constant, $T_{\rm fl,p}$ is the floor temperature motivated by the observations \citep[e.g.,][]{Inserra2013,Nicholl2017,Omand2024}, and we approximate the photospheric velocity as $v_{\rm ph,p} \simeq v_{\rm ej,p}$, {which is a standard reasonable assumption in the literature \citep[e.g.,][]{Inserra2013,Nicholl2017}}. For the magnetar-powered emission from the near-equatorial SN ejecta and evaporated material $L_{\rm mag,e}$, one can calculate the photosphere temperature $T_{\rm ph,e}$ by replacing $L_{\rm mag,p}$, $T_{\rm fl,p}$, $v_{\rm ph,p}$, and the factor $1-f_{\rm ev}$ in Equation (\ref{equ:Temperature}) with $L_{\rm mag,e}$, $T_{\rm fl,e}$, $v_{\rm ph,e} \simeq  v_{\rm ej,e}$, and $f_{\rm ev}$, respectively. Besides the floor temperature,  extinction from the Milky Way and the SLSN host galaxy can affect the color evolution of lightcurve. We take a fixed value of $R_V=3.1$ for the Milky Way extinction \citep{Schlafly2011}, while the host extinction $A_V$ is assumed to be a free parameter.

We infer the free parameters in the magnetar-star binary engine model for the multiband lightcurves of the collected bumpy SLSNe with a Markov Chain Monte Carlo method, using the \texttt{emcee} package \citep{ForemanMackey2013}. By setting the first observed data as the zero point, we define a time $t_{\rm shift}$ to leftward shift SLSN lightcurve. In total, our model includes 12 free parameters. The log likelihood function is $\ln\mathcal{L} = -0.5\sum_{i=1}^n(O_i-M_i)^2/\sigma_i^2$, where $O_i$, $\sigma_i$, and $M_i$ are the $i$-th of $n$ observed apparent magnitudes, observation uncertainties, and model apparent magnitudes, respectively. The priors on these fitting parameters, which are assumed to be independent, and posteriors medians with $16\%$ and $84\%$ quantiles for each event are listed in Table \ref{tab:PriorsAndResults}. The best fits to the multiband lightcurves are presented in Figure \ref{fig:Fitting_Results}. An example corner plot of the posterior probability distributions is shown in Figure \ref{fig:Posterior} in Appendix \ref{sec:Corner}.

Generally, the primary peaks of the multiband SLSN lightcurves, as well as the post-peak bumps, can be well fit by our magnetar-star binary engine model.  In previous literature, the typical values of the magnetar parameters and ejecta mass were given as $P_{\rm i}\sim3\,{\rm ms}$, $B_{\rm p}\sim2\times10^{14}\,{\rm G}$, and $M_{\rm ej}\sim4\,M_\odot$ \citep[e.g.,][]{Yu2017,Liu2017,Nicholl2017,Kumar2024}. As shown in Table \ref{tab:PriorsAndResults}, our inferred values of the magnetar parameters and ejecta mass are $P_{\rm i}\sim2-4\,{\rm ms}$, $B_{\rm p}\sim6\times10^{13}-1.5\times10^{14}\,{\rm G}$, and $M_{\rm ej}\sim2.5-6\,M_\odot$, which are consistent with the statistical results presented in the literature. The masses of the evaporated material in our bumpy SLSN sample are concentrated around $\sim0.5\,M_\odot$, implying that only a fraction of the material from the companion stars can be stripped and evaporated by the newborn magnetars. PTF10hgi, whose post-peak bump is the least significant in our sample, has the lowest $f_{\rm ev}$ at $\sim0.24$, while $f_{\rm ev}$ for other collected SLSNe can be as high as $\sim0.5-0.6$.

\section{Summary and Discussion}
\label{sec:DiscussionConclusion}

The effect of companions on SN lightcurves has already been discussed in the literature \citep[e.g.,][]{Soker2020,Gao2020,Hirai2022,Wen2023}, and has been observed in SN2022jli \citep{Moore2023,Chen2024}, identified as a normal Type Ic SN. In this {\em Letter}, we propose a magnetar-star binary engine model to explain bumpy SLSNe. Our model can accurately reproduce the primary peak and post-peak bump of the collected multiband lightcurves of bumpy SLSNe. In the following, we suggest that this magnetar-star binary engine can offer an explanation for the polarization evolution, as well as for the origin of velocity broadening of late-time broad hydrogen or helium features, observed in some bumpy SLSNe. It is expected that the diversity in the lightcurves and spectra of SLSNe may be attributed to the wide variety of companion stars and post-SN binary systems.

\subsection{Origin of broad-line H$\alpha$ lines}

\cite{Yan2015,Yan2017} reported three bumpy SLSNe that exhibit broad-line H$\alpha$ emission with velocity widths of $\sim4000-6000\,{\rm km}\,{\rm s}^{-1}$ at $>100\,{\rm d}$ after the peaks. They also proposed that $\sim10-30\%$ of SLSNe could display such hydrogen features in their spectra. \cite{Quimby2018} identified that PTF10hgi, which is also discussed in this {\em Letter} as one of our fitted SLSNe, shows late-time H$\alpha$ with a velocity of $10,000\,{\rm km}\,{\rm s}^{-1}$. Furthermore, these observed H$\alpha$ lines have an overall blue-ward shift relative to the SLSN hosts. In the literature, some SLSN bumps and these late-time H$\alpha$ emissions are usually attributed to the ejecta colliding with a hydrogen-rich CSM \citep[e.g.,][]{Yan2015,Yan2017,Nicholl2016,Wang2016,Liu2018}. It was also proposed that H$\alpha$ emission could originate from material stripped off the MS companion by the SN explosion \citep[e.g.,][]{Moriya2015,Hirai2018}. However, simulations by \cite{Liu2015} reached the opposite conclusion: that most SNe would lead to inefficient mass stripping of their MS companion stars.

Our magnetar-star binary engine model can naturally explain the origin of broad-line H$\alpha$ lines from bumpy SLSNe. Since the evaporated material is distributed in the inner layer of the ejecta, the MW can excite this hydrogen-rich material, which can be observable once the ejecta become transparent. Following the fitting results of our SLSN sample listed in Table \ref{tab:PriorsAndResults}, the velocity of the evaporated material $v_{\rm ej,e}$ can range from $\sim6000$ to $13,000\,{\rm km}\,{\rm s}^{-1}$. By assuming that $v_{\rm ej,e}$ is equal to the velocity of H$\alpha$ line broadening, our inferred velocities are roughly consistent with the observation values \citep{Yan2015,Yan2017,Quimby2018}. The overall blueshift of the observed H$\alpha$ emission could be caused by a viewing angle effect due to a potentially complicated angular distribution of the evaporated material.

Population synthesis simulations by \cite{Hu2023} indicate that $\sim65-75\%$ of companions of SLSN progenitors in close binaries are MS stars, while $\sim10-15\%$ are majorly $\lesssim1-2\,M_\odot$ helium stars. Consequently, one might expect to detect broad helium features in the spectra of some SLSNe, as confirmed by \cite{Yan2020}.

\subsection{Origin of Multiple Post-peak Bumps and Diversity of SLSN Lightcurves}

{\cite{Hu2023} suggested that SLSN progenitors could originate from $\sim5-40\,M_\odot$ Wolf-Rayet stars in close binary systems with an orbital period of $\lesssim2\,{\rm d}$. Systems with companions more massive than the Wolf-Rayet star in close binaries are more likely to remain bound after the supernova explosion \citep[e.g.,][]{Tauris2017}. The wide variety of post-SN binary systems could lead to a diversity of bumps in SLSN lightcurves.}

This {\em Letter} suggests that SLSN lightcurves with one post-peak bump can result from the evaporation of companion stars by newborn magnetars when the post-SN binary systems have a close circular orbit. If the binary is unbound by the SN explosion, reaching an eccentricity  $e\geq1$, but the magnetar can skim close to the surface of the MS companion, the MW could evaporate a large amount of material at the pericenter. As a result, the final declining SLSN lightcurve could also have only one bump. 

{There are also many SLSNe with observed lightcurves showing multiple post-peak bumps, e.g., SN2015bn \citep{Nicholl2016}, SN2017egm \citep{Zhu2023}, and SN2020qlb \citep{West2023}.} Due to the abrupt mass loss and natal kick accompanying a SN explosion, the post-SN binary systems may remain bound with very high eccentricities. Newborn magnetars could induce violent evaporation from their companion stars near the perihelions periodically, resulting in SLSN lightcurves with multiple bumps. This actually provides a plausible physical explanation for the flare activity supposed in \cite{Yu2017Flare}. Observationally, $\sim20-70\%$ SLSNe can be well explained by a smooth magnetar model \citep[e.g.,][]{Hosseinzadeh2022,Chen2023ii}. One possible reason is that these systems are directly disrupted by {abrupt mass loss and} SN kicks, and the magnetars and their companions move in opposite directions, resulting in very little evaporated material. Additionally, \cite{Hu2023} found via population synthesis modelling that $\sim10-20\%$ of companions of SLSN progenitors could be NSs or BHs. Although the systems are not disrupted, the final SLSN lightcurves can plausibly be powered by a magnetar engine alone.

\subsection{Viewing Angle Effect}

If the post-SN orbit is circular, the evaporated material typically distributes around the equatorial plane. The magnetar-powered emission in the polar direction is expected to be brighter than that in the equatorial direction. Thus, the SLSN emission powered by the magnetar-star binary engine should be viewing-angle-dependent. Furthermore, if the post-SN orbit is highly eccentric, the evaporated material might be even more anisotropic, acquiring a shape possibly resembling a crescent, which could lead to a greater viewing angle effect on the SLSN lightcurves.

The asymmetric explosion of SLSNe powered by the magnetar-star binary engine could yield polarized emission. Shortly after the explosion, the central injection from the magnetar has little influence on the nearly-symmetric outer ejecta. The photosphere is at the outer edge of the ejecta, resulting in low observed polarization. Over time, the magnetar central injection causes the polar ejecta, with lower mass {per solid angle}, to move faster, while the equatorial ejecta and evaporated material, with higher mass {per solid angle}, move more slowly. As indicated in Figure \ref{fig:Cartoon}, the final geometrical structure of the material at the inner boundary could be ellipsoidal. As the photosphere penetrates deeper inside the ejecta and evaporated material, an increase in polarization will be observed. This polarization evolution and ejecta profile has been confirmed for some bumpy SLSNe, e.g., SN2015bn and SN2017egm \citep[e.g.,][]{Inserra2016,Leloudas2017,Saito2020}.

The best-fit values of $f_{\rm ev}$ in our sample reach as high as $\sim0.66$, corresponding to $\theta_{\rm ev}\sim0.72$, which would exceed the angular radius of the companion star as viewed from the magnetar, i.e., $\theta_{\star}\sim R_\star/a\sim0.23\,R_{\star,3\odot}a^{-1}_{13\odot}$. One possible reason is that as the evaporated material moves outward, it could spread over a greater angle, potentially causing $\theta_{\rm ev}$ to be larger than $\theta_{\star}$. Additionally, the line of sight of some SLSNe may be close equatorial, resulting in a larger observed proportion of magnetar-powered emission from the equatorial ejecta and evaporated material. More detailed dynamical evolution of the evaporated material, as well as more comprehensive radiation transport simulations of the viewing-angle-dependent lightcurve and polarization evolution of SLSN emissions powered by the magnetar-star binary engine, are topics for further study.

\software{\texttt{Python}, \url{https://www.python.org}; \texttt{Matlab}, \url{https://www.mathworks.com}; \texttt{emcee} \citep{ForemanMackey2013}; \texttt{corner} \citep{ForemanMackey2016}; \texttt{OriginPro}, \url{https://www.originlab.com/originpro}; \texttt{Procreate}, \url{https://procreate.com}}

\acknowledgments {The authors acknowledge an anonymous referee for useful discussions.} We thank Ping Chen, Zi-Gao Dai, Jim Fuller, He Gao, Rui-Chong Hu, Yachen Kang, Ying Qin, Yong Shao, Guang-Lei Wu, and Yuan-Pei Yang for helpful comments. JPZ, IM \& RH acknowledge support from the Australian Research Council Centre of Excellence for Gravitational Wave Discovery (OzGrav), through project number CE17010004. IM is a recipient of the Australian Research Council Future Fellowship FT190100574. LDL is supported by the National Natural Science Foundation of China under Grant No. 12303047 and the Hubei Provincial Natural Science Foundation under Grant No. 2023AFB321. YWY is supported by the science research grant from the China Manned Space Project with No. CMS-CSST-2021-A12, the National Natural Science Foundation of China (Grant No. 12393811), the National SKA Program of China (2020SKA0120300), and the National Key Research and Development Programs of China (2021YFA0718500). AC is supported by the China Postdoctoral Science Foundation (No. 2023T160410). 
\appendix

\renewcommand{\thesection}{\Alph{section}}

\section{Emission from Radioactive Decay and Interaction of Evaporated Material with Ejecta}
\label{sec:OtherEmission}

\renewcommand{\theequation}{A\arabic{equation}}

The injection luminosity powered by the energy released via the radioactive decay of $^{56}$Ni and $^{56}$Co is given by \citep{Nadyozhin1994,Arnett1996}
\begin{equation}
    L_{\rm r} = (6.54e^{-t/t_{\rm Ni}} + 1.45e^{-t/t_{\rm Co}})\frac{M_{\rm Ni}}{1\,M_\odot}\times10^{43}{\rm erg}\,{\rm s}^{-1},
\end{equation}
where $t_{\rm Ni}=8.8\,{\rm d}$ and $t_{\rm Co} = 111.3\,{\rm d}$ are the mean lifetimes of the nuclei, and $M_{\rm {Ni}}$ is the nickel mass in solar masses, which we scale to $M_{\rm Ni}=0.1M_{\rm ej}$. Thus, the $^{56}$Ni-powered bolometric lightcurve from the polar and equatorial directions can be expressed as
\begin{equation}
\begin{split}
    L_{\rm rad,p}(t) &= (1-f_{\rm ev})e^{-(t/t_{\rm d,p})^2}\int_0^{t}2L_{\rm r}(t')\frac{t'}{t_{\rm d,p}}e^{(t'/t_{\rm d,p})^2}\frac{dt'}{t_{\rm d,p}}, \\
    L_{\rm rad,e}(t) &= f_{\rm ev}e^{-(t/t^{\rm rad}_{\rm d,e})^2}\int_0^{t}2L_{\rm r}(t')\frac{t'}{t^{\rm rad}_{\rm d,e}}e^{(t'/t^{\rm rad}_{\rm d,e})^2}\frac{dt'}{t^{\rm rad}_{\rm d,e}},
\end{split}
\end{equation}
where we note the effective diffusion time from the equatorial direction is $t_{\rm d,e}^{{\rm rad}} = (3\kappa_{\rm ej}M_{\rm ej}/4\pi{v}_{\rm ej,e}c)^{1/2}$ since $^{56}$Ni is located at the inner part of the ejecta. 

\begin{figure}[t!]
\centering
\includegraphics[width=1\linewidth, trim = 70 36 94 60, clip]{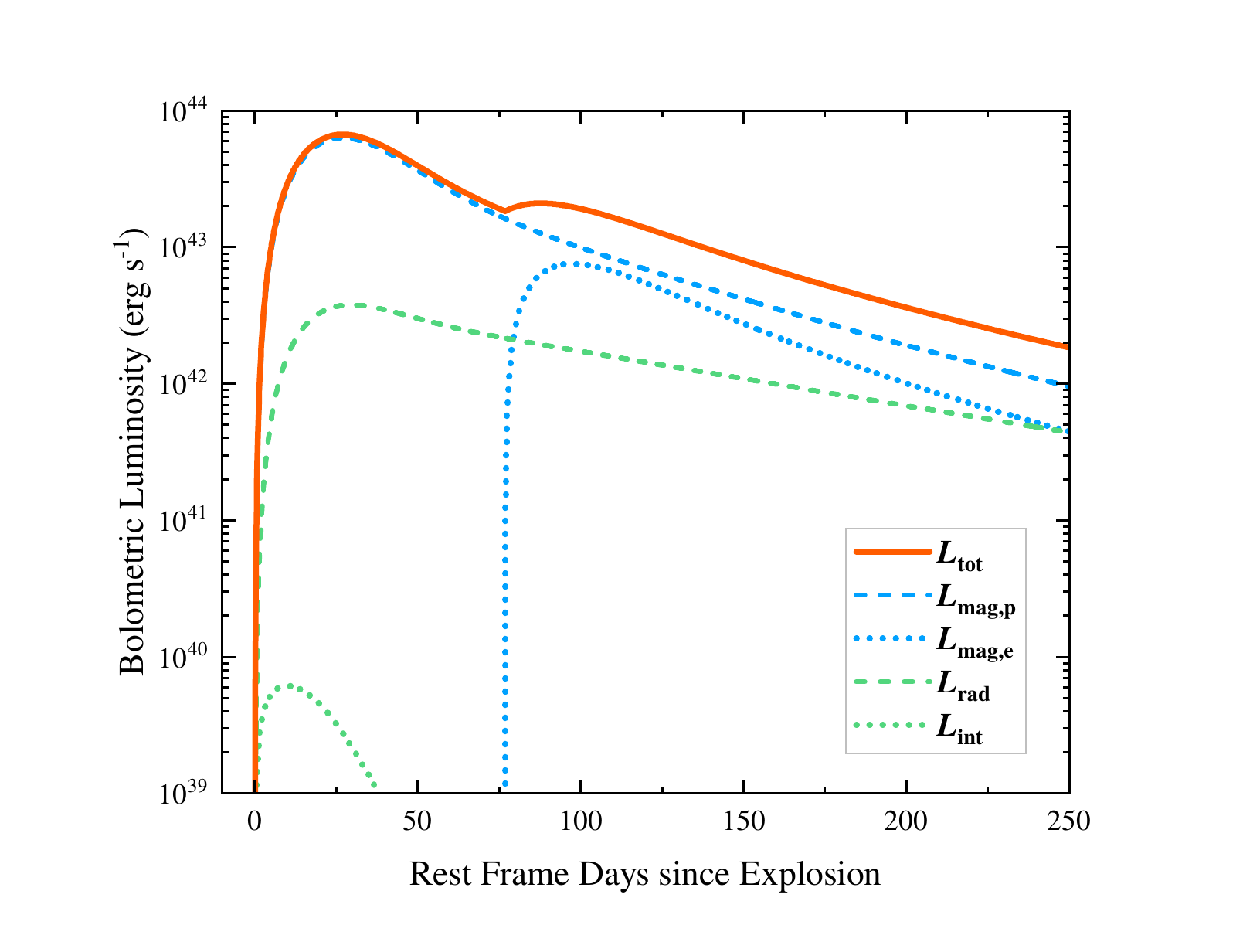}
\caption{Bolometric lightcurves based on the best fit to the multiband lightcurves of SN2018kyt. The blue dashed, blue dotted, green dashed, and green dotted lines represent the contributions from the magnetar injection from the polar ejecta, the magnetar injection from the equatorial evaporated material and ejecta, radioactive heating, and interaction between the  evaporated material and ejecta, respectively. The solid red line is the total luminosity.}
\label{fig:Contribution}
\end{figure}

After the material evaporated from the MS companion is accelerated by the MW, this material can collide with and interact with the ejecta. Because the evaporated material can block the MW, preventing the acceleration of the equatorial SN ejecta, we assume that the ejecta have a broken power-law density profile following SN numerical simulations \citep{Chevalier1989,Matzner1999}:
\begin{equation}
    \rho_{\rm SN}(v,t) = \left\{\begin{aligned}
 \zeta_\rho\frac{M_{\rm ej}}{v_{\rm tr}^3t^3}\left(\frac{v}{v_{\rm tr}} \right)^{-\delta},~v<v_{\rm tr}, \\
 \zeta_\rho\frac{M_{\rm ej}}{v_{\rm tr}^3t^3}\left(\frac{v}{v_{\rm tr}} \right)^{-n},~v\geq v_{\rm tr},
\end{aligned}\right.
\end{equation}
with a transition of the density occurring at a velocity coordinate of $v_{\rm tr}=\zeta_\nu\sqrt{{E_{\rm SN}}/{M_{\rm ej}}}$, where $\zeta_\rho = (n-3)(3-\delta)/4\pi(n-\delta)$ and $\zeta_\nu = \sqrt{2(5-\delta)(n-5)/(n-3)(3-\delta)}$. For core-collapse SNe, we adopt typical values of $n=10$ and $\delta=1$ as fiducial \citep{Chevalier1989}. The high-pressure MW can sweep the evaporated material into a thin shell. Since the shell moves faster than the local ejecta, a radiation-dominated shock can be formed at the front of the shell. The conservation equations of mass, momentum, and energy  describe the shell dynamics \citep[e.g.,][]{Chevalier1992,chevalier2005,Kasen2016,Liu2021}:
\begin{equation}
\begin{split}
\label{equ:DynamcialEquation}
    \frac{dM_{\rm sh}}{dt} &= 4\pi R_{\rm sh}^2\rho_{\rm SN}(v_{\rm sh}-v_{\rm SN}), \\
    M_{\rm sh}\frac{dv_{\rm sh}}{dt} &= 4\pi R_{\rm sh}^2\left[p_{\rm b} - \rho_{\rm SN}(v_{\rm sh} - v_{\rm SN})^2\right],\\
    \frac{d(4\pi{p}_{\rm b}R_{\rm sh}^3)}{dt} &= L_{\rm sd}(t) - 4\pi{p}_{\rm b}R_{\rm sh}^2\frac{dR_{\rm sh}}{dt} - \frac{L_{\rm mag,e}(t)}{f_{\rm ev}},
\end{split}
\end{equation}
where the shell mass, radius, and velocity are $M_{\rm sh}$, $R_{\rm sh}$, $v_{\rm sh}=dR_{\rm sh}/dt$, respectively, $v_{\rm SN}=R_{\rm sh}/t$ is the ejecta velocity at radius $R_{\rm sh}$, $\rho_{\rm SN}$ is the pre-shock ejecta density at the front of the shell, and $p_{\rm b}$ is the pressure in the magnetar-driven bubble. For simplicity, we assume the initial values of $M_{\rm sh}$ and $v_{\rm sh}$ are $M_{\rm ev}/f_{\rm ev}$ and $v_{\rm esc}$. The local heating rate of the shock is given by \citep{Kasen2016,Li2016,Liu2021}
\begin{equation}
    H_{\rm sh}(t) = \frac{1}{2}\frac{dM_{\rm sh}}{dt}(v_{\rm sh} - v_{\rm SN})^2.
\end{equation}
The interaction emission can be modelled by the standard SN diffusion equation \citep{Arnett1982,Liu2021} as
\begin{equation}
    L_{\rm int}(t) = f_{\rm ev}e^{-(t/t_{\rm bo})^2}\int^{t}_02H_{\rm sh}(t')\frac{t'}{t_{\rm bo}}e^{(t'/t_{\rm bo})^2}\frac{dt'}{t_{\rm bo}},
\end{equation}
where the breakout time of the photons at the shock $t_{\rm bo}$ can be estimated by setting the diffusion time equal to the shock propagation time, i.e., solving the equation $\tau_{\rm SN} \equiv c/v_{\rm sh} = \int_{v_{\rm SN}}^{+\infty}\kappa_{\rm ej}\rho_{\rm SN}(v,t_{\rm bo})t_{\rm bo}dv$. 

As an example, we show the contributions of different components to a SLSN lightcurve in the magnetar-star binary engine based on our best-fit parameters for SN2018kyt in Figure \ref{fig:Contribution}. The $^{56}$Ni-powered luminosity is about an order of magnitude lower than the magnetar-powered luminosity. Meanwhile, interaction contributes less than $10^{-4}$ of the magnetar-powered luminosity. We note that our model might overestimate the peak luminosity and underestimate the peak timescale of the interaction emission, since our simplified model assumes that the initial mass of the evaporated material is $M_{\rm ev}$ in Equation (\ref{equ:DynamcialEquation}) while the real evaporation process is expected to take time. Thus, it is safe to neglect the contributions from radioactive decay and interaction.

\section{Posterior Probability Distributions} \label{sec:Corner}

\begin{figure*}
\centering
\includegraphics[width=1\linewidth]{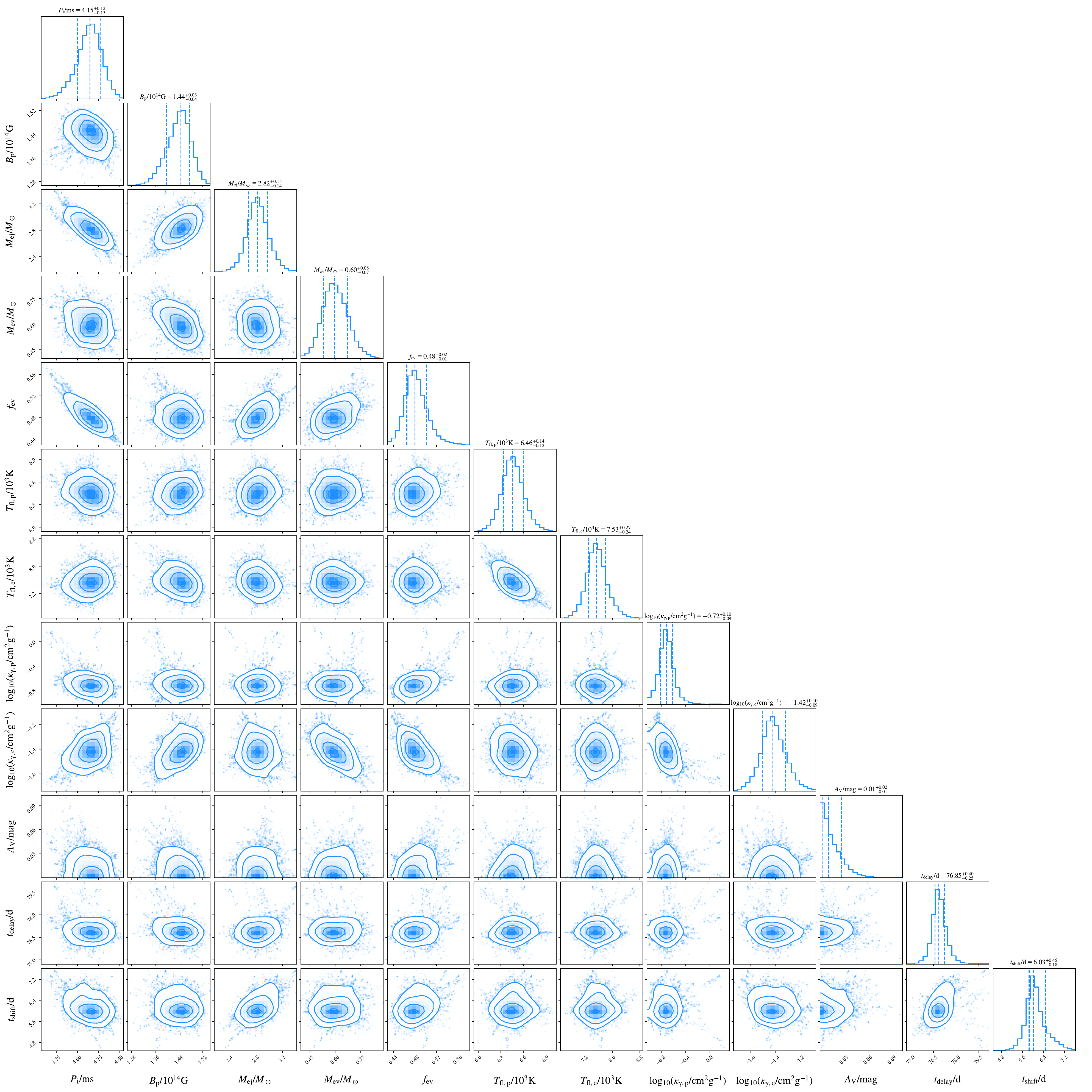}
\caption{Posteriors on the fit parameters for SN2018kyt. Medians with $16\%$ and $84\%$ quantiles are labeled.}
\label{fig:Posterior}
\end{figure*}

For the convenience of readers in understanding the convergence of our fits, the posterior distributions of the fit parameters for SN2019stc are illustrated in Figure \ref{fig:Posterior} as an example.

\clearpage

\bibliography{ms}{}
\bibliographystyle{aasjournal}
\end{document}